\documentclass[conference]{IEEEtran}

\usepackage{amsmath,amssymb,amsfonts,bm,mathtools}
\usepackage{graphicx}
\usepackage{xcolor}
\usepackage{cite}
\usepackage{booktabs}
\usepackage{multirow}
\usepackage{array}
\usepackage{url}
\begin{document}

\title{Active Beyond-Diagonal RIS Empowered Heterogeneous Edge Computing: A Distributional Reinforcement Learning Approach}

\author{
\IEEEauthorblockN{
Tianyu Pang\IEEEauthorrefmark{1} and
Hongyu Li\IEEEauthorrefmark{2}
}
\IEEEauthorblockA{
Internet of Things Thrust, The Hong Kong University of Science and Technology (Guangzhou), Guangzhou, China\\
\IEEEauthorrefmark{1}E-mail: \texttt{tpang457@connect.hkust-gz.edu.cn};
\IEEEauthorrefmark{2}E-mail: \texttt{hongyuli@hkust-gz.edu.cn}
}
}

\maketitle

\begin{abstract}
Active beyond-diagonal reconfigurable intelligent surfaces (BD-RISs) enables hybrid transmitting and reflecting mode to achieve effective signal amplification and full-space coverage, thus providing a promising solution for blockage-aware uplink offloading in heterogeneous mobile edge computing (MEC) systems. 
However, practical hybrid mode active BD-RIS are realized by reciprocal devices, which inherently generate cross-sector energy leakage that will reshape the system-level energy-latency tradeoff.
This paper studies energy-aware offloading and resource allocation for reciprocal active BD-RIS-assisted heterogeneous MEC, where offloading decisions, CPU/GPU computation allocation, transmit powers, receive processing, and active BD-RIS are tightly coupled. The resulting problem is a high-dimensional mixed integer nonconvex problem and is difficult to solve efficiently by conventional per-instance optimization. 
To address this challenge, we develop an end-to-end joint optimization framework based on a refined version of the distributional soft actor--critic algorithm, named as DSAC-T.
By modeling return distributions rather than only expected values, DSAC-T improves policy stability under reward heterogeneity and feasibility-boundary sensitivity. 
Compared with other baseline algorithms, DSAC-T achieves the best energy-latency reward, the highest feasibility ratio of $81.67$\%, and a fast online decision time of $0.0267$ s per scenario.
\end{abstract}

\begin{IEEEkeywords}
Active, BD-RIS, DSAC-T, Heterogeneous MEC, Deep Reinforcement Learning.
\end{IEEEkeywords}

\section{Introduction}

Mobile edge computing (MEC) enables latency-sensitive and computation-intensive services by offloading demanding workloads to nearby edge servers \cite{Mao2017Survey}. In practical edge intelligence scenarios, task demands are heterogeneous, since CPU- and GPU-oriented jobs coexist and compete for different edge-side resources. Therefore, task offloading is inherently a communication--computation co-design problem. 
Wireless communications are very often interrupted by obstacles, such as high buildings, while reconfigurable intelligent surfaces (RISs) \cite{liu2021reconfigurable} have been proven to provide a flexible way to enhance wireless connectivity and thus MEC \cite{Hu2021RISMECOptimizationLearning}.

The RIS technology has grown rapidly with the emergence of many advanced architectures, which can be broadly categorized into passive RISs and active RISs based on if or not the scattered signals can be amplified.
In the passive form, beyond-diagonal (BD) RISs \cite{Li2026Tutorial} are a universal and generalized framework that includes conventional passive RISs with diagonal phase shift matrices \cite{liu2021reconfigurable} and simultaneous transmitting and reflecting (STAR) RISs \cite{Liu2021STAR360Coverage} as special cases, while introducing inter-element connections to open the door for supporting more advanced hybrid and multi-sector modes with more flexible wave manipulation and wider coverage. 
These benefits have motivated the integration of BD-RIS and MEC to improve computing performance \cite{Qin2025Joint}. 

Despite this progress, passive RISs have fundamental performance limits due to the multiplicative fading. Hence, active RISs with the ability to amplify signals using reflection-type amplifiers have been proposed to compensate for the multiplicative fading and further enhance wireless channels \cite{Long2021ActiveRISAided,Xu2023Active}. Inspired by this point, BD-RIS has been recently upgraded to the active form, with more rigorous physics-consistent model and diverse architecture design \cite{shen2026active}, and more advanced hybrid mode \cite{liu2026active} that includes active STAR-RIS \cite{Xu2023Active} as a special case. 

In the family of active BD-RIS, there have been studies on active STAR-RIS-assisted MEC \cite{aung2025active,qin2025resource}, while the active STAR-RIS models used do not fully expose the coupling between transmitting sector and reflecting sector due to the inherent circuit reciprocity.   
This essentially means when blocked users are assisted through active RIS across sectors, the system incurs active forwarding cost and cross-sector energy leakage due to the reciprocity, which compresses the energy budget available for uplink transmission and heterogeneous edge execution. Thus, simply enhancing blocked-user connectivity may improve communication reliability but degrade overall energy efficiency or aggravate latency violation.
Moreover, the aforementioned RIS-assisted MEC works use generic edge-computing abstractions without explicitly distinguishing CPU/GPU-oriented tasks or decoupled CPU/GPU edge resource pools. Deeply blocked direct links are also often omitted or over-idealized, whereas they are more appropriately modeled as severely attenuated residual paths in practice. These limitations prevent existing formulations from fully capturing the real-world heterogeneous MEC systems.

Motivated by these observations, we study a physics-consistent active BD-RIS-assisted heterogeneous MEC system.

The contributions are summarized as follows. 

\textit{First}, we establish an active BD-RIS-assisted heterogeneous MEC model that captures asymmetric blockage, residual direct-link attenuation, active RIS power consumption, amplified noise, and decoupled CPU/GPU edge resources. 
Specifically, the active BD-RIS works on the hybrid mode (similar to active STAR-RIS) to support full-space coverage and, more importantly, is modeled rigorously based on a reciprocal circuit implementation to capture cross-sector leakage \cite{liu2026active}. 

\textit{Second}, we formulate and solve a mixed integer, high-dimensional, and strongly coupled nonconvex optimization problem that jointly involves offloading decisions, computation allocation, uplink transmit powers, receive processing, and the active BD-RIS. 
Deep reinforcement learning (DRL) offers an efficient way to solve such problems \cite{Huang2020DRLOnlineOffloadingMEC,Peng2023RobustDRLForEHRIS}, while the formulated system remains challenging for conventional expectation-based actor--critic methods. Blockage patterns, task heterogeneity, active RIS cost, cross-sector leakage, and latency penalties jointly induce strong reward heterogeneity and feasibility-boundary sensitivity, leading to unstable value estimation and policy updates. 
To address this issue, we adopt a refined version of the distributional soft actor--critic (DSAC) algorithm that is suitable for solving problems with highly heterogeneous rewards, namely DSAC-T \cite{Duan2025DSACT}, and develop an end-to-end DSAC-T-based joint optimization framework. 


\textit{Third}, simulation results validate the effectiveness and efficiency of the proposed DSAC-T-based framework. Over 300 evaluation scenarios, DSAC-T achieves the best energy-latency reward and the highest feasibility ratio of $81.67\%$, outperforming other baselines. It also requires only $0.0267$ s per scenario for online decision making, which is orders of magnitude faster than the mathematical optimization algorithm.

\section{System Model and Problem Formulation}

In this section, we present the active BD-RIS assisted heterogeneous MEC system, including the blockage-aware channel model, the uplink transmission process, the heterogeneous local/edge computation model, and the joint offloading and resource-allocation problem.

\begin{figure}[t]
    \centering
    \includegraphics[width=0.95\columnwidth]{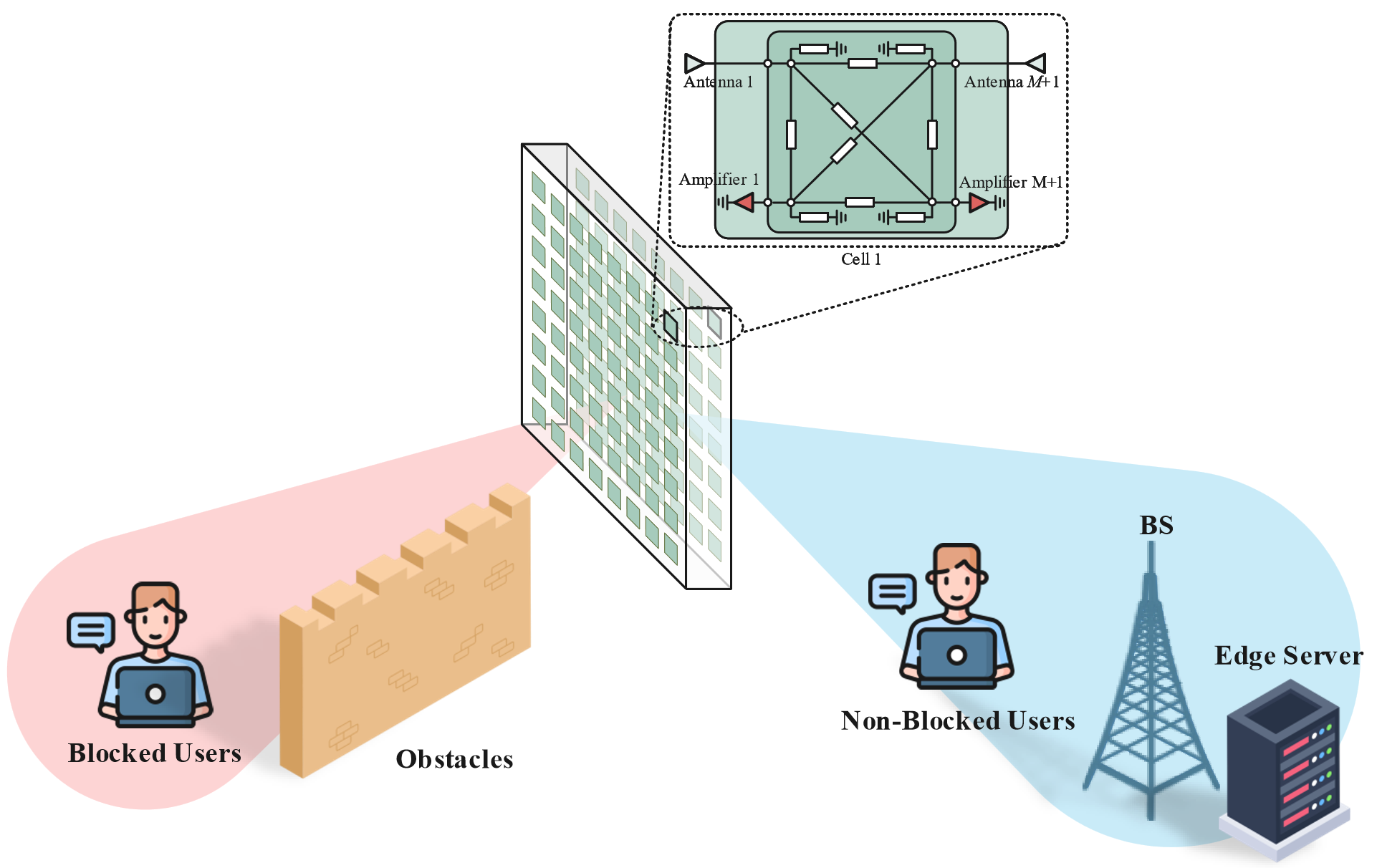}
    \caption{Diagram of an active BD-RIS assisted heterogeneous MEC system.}
    \label{fig:system_model}
\end{figure}

\subsection{Active BD-RIS Assisted Channel Model}

We consider an uplink heterogeneous MEC system consisting of one edge server equipped with an $N$-antenna base station (BS), one $M$-cell hybrid mode active BD-RIS \cite{liu2026active}, and $K$ single-antenna users. Each cell of the hybrid mode active BD-RIS is constructed by two back to back placed antenna elements connected to a 2-port active reconfigurable impedance network, which further consists of two reflection-type amplifiers connected to a 4-port passive reconfigurable impedance network. In this sense, the hybrid mode active BD-RIS has two sectors, namely transmitting sector (sector 2) and reflecting sector (sector 1), each of which contains $M$ elements and covers half of the space, as illustrated in Fig. ~\ref{fig:system_model}.  Therefore, the BS is located within the reflecting sector of the hybrid mode active BD-RIS. In addition, the user set is partitioned into two disjoint subsets, denoted by $\mathcal{K}_1$ with $|\mathcal{K}_1|=K_1$ and $\mathcal{K}_2$ with $|\mathcal{K}_2|=K_2$, $K_1 + K_2 = K$, corresponding to non-blocked users located within the reflecting sector and deeply blocked users within the transmitting sector of hybrid mode active BD-RIS, respectively. Each user is associated with either a CPU-oriented task or a GPU-oriented task. The edge server maintains separate CPU and GPU resource pools to support heterogeneous task execution.

\subsubsection{Direct Channel} The direct channel from user $k$ to BS is modeled as $\mathbf{h}_{\mathrm{d},k} = \beta_k\tilde{\mathbf{h}}_{\mathrm{d},k}\in\mathbb{C}^{N\times 1}$, where $\beta_k = 1$ if $k\in\mathcal{K}_1$ and $\beta_k \ll 1$ if $k\in\mathcal{K}_2$. 
Hence, for users in $\mathcal{K}_2$, the direct link is approximately zero but not strictly removed.


\subsubsection{Hybrid Mode Active BD-RIS} The $M$-cell hybrid mode active BD-RIS is characterized by its scattering matrix \cite{liu2026active}
\begin{equation}
    \mathbf{\Theta} =\left[ \begin{matrix}\mathbf{\Theta}_{1,1} &\mathbf{\Theta}_{1,2}\\
    \mathbf{\Theta}_{2,1} &\mathbf{\Theta}_{2,2}\end{matrix}\right],
\end{equation}
where $\mathbf{\Theta}_{i,j} \in\mathbb{C}^{M\times M}$. In this work, we assume the hybrid mode active BD-RIS has a reciprocal cell-wise single-connected architecture, such that $\mathbf{\Theta} = \mathbf{\Theta}^\mathsf{T}$ and $\mathbf{\Theta}_{i,j}$, $\forall i,j$ are diagonal, further implying that $\mathbf{\Theta}_{1,2} = \mathbf{\Theta}_{2,1}$.  

\subsubsection{Active BD-RIS Assisted Channel} Let $\mathbf{H}\in\mathbb{C}^{N\times M}$ denote the RIS-BS channel and $\mathbf{h}_{\mathrm{r},k}\in\mathbb{C}^{M\times 1}$ denote the user-RIS channel of user $k$. 
The effective uplink channel of user $k$ is then given by
\begin{equation}
\mathbf{h}_{k}
= \mathbf{h}_{\mathrm{d},k} + \mathbf{H}\mathbf{\Theta}_{1,i}\mathbf{h}_{\mathrm{r},k},\forall k\in\mathcal{K}_i,\forall i\in\{1,2\}.
\end{equation}

\subsection{Uplink Transmission Model}

Let $p_k$ denote the uplink transmit power of user $k$ and $\mathbf{w}_k$ denote the combiner associated with user $k$. The uplink signal-to-interference-plus-noise ratio (SINR) can be expressed as
\begin{equation}
\gamma_k
=
\frac{p_k\left|\mathbf{w}_k^\mathsf{H}\mathbf{h}_{k}\right|^2}
{\sum\limits_{j\neq k}p_j\left|\mathbf{w}_\mathsf{k}^\mathsf{H}\mathbf{h}_{j}\right|^2
+\sigma_\mathrm{B}^2\|\mathbf{w}_k\|_2^2
+\sigma_\mathrm{I}^2\left\|\mathbf{w}_k^\mathsf{H}\mathbf{H}\bar{\mathbf{\Theta}}\right\|_2^2},
\end{equation}
where $\bar{\mathbf{\Theta}} = [\mathbf{\Theta}_{1,1}~\mathbf{\Theta}_{1,2}]$,  $\sigma_\mathrm{B}^2$ denote the noise power at the BS and $\sigma_\mathrm{I}^2$ denote the dynamic noise power of active BD-RIS.
The achievable uplink rate of user $k$ is thus 
$R_k = B\log_2(1+\gamma_k)$, where $B$ is the system bandwidth.




\subsection{Heterogeneous Computation Model}

The task of user $k$ is characterized by the tuple
$\left(\tau_k, D_k, C_k, T_k^{\max}\right)$,
where $\tau_k\in\{\mathrm{CPU},\mathrm{GPU}\}$ denotes the task type, $D_k$ is the input data size, $C_k$ is the required CPU cycles per bit, and $T_k^{\max}$ is the latency deadline.
Let $\alpha_k\in\{0,1\}$ denote the offloading decision, where $\alpha_k=0$ indicates local execution and $\alpha_k=1$ indicates edge execution. The local execution latency and offloading latency that involves transmission and edge execution components are respectively 
\begin{equation}
T_k^{\mathrm{loc}}=\frac{D_k C_k}{f_k^{\mathrm{loc}}},~T_k^{\mathrm{off}}=\frac{D_k}{R_k} + \frac{D_k C_k}{f_k^{\mathrm{edge}}},
\end{equation}
where $f_k^{\mathrm{loc}}$ is the local computing rate and $f_k^\mathrm{edge}$ is the edge computing rate allocated to user $k$.
Accordingly, the end-to-end latency is constrained by 
\begin{equation}
T_k=(1-\alpha_k)T_k^{\mathrm{loc}}+\alpha_kT_k^{\mathrm{off}}\le T_k^\mathrm{max}.
\end{equation}

To capture hardware-specific execution, the edge computation budget is decoupled into CPU and GPU resource pools. Let $F_{\mathrm{cpu}}^{\mathrm{edge}}$ and $F_{\mathrm{gpu}}^{\mathrm{edge}}$ denote the total edge CPU and GPU capacities, respectively. Then,
\begin{equation}
\sum_{k\in\mathcal{K}^{\mathrm{cpu,off}}} f_k^{\mathrm{edge}}\leq F_{\mathrm{cpu}}^{\mathrm{edge}},\;
\sum_{k\in\mathcal{K}^{\mathrm{gpu,off}}} f_k^{\mathrm{edge}}\leq F_{\mathrm{gpu}}^{\mathrm{edge}},
\end{equation}
where $\mathcal{K}^{\mathrm{cpu,off}}$ and $\mathcal{K}^{\mathrm{gpu,off}}$ denote the sets of offloaded CPU-type and GPU-type tasks, respectively.

\subsection{Energy Consumption Model}

The power radiated from active BD-RIS is
\begin{equation}
P_\mathrm{RIS} = \sum_{k=1}^Kp_k\left\|\mathbf{\Theta}\bar{\mathbf{h}}_{\mathrm{r},k}\right\|_2^2
+ \sigma_\mathrm{I}^2\left\|\mathbf{\Theta}\right\|_\mathsf{F}^2\le P_\mathrm{A},
\end{equation}
where $\bar{\mathbf{h}}_{\mathrm{r},k} = [\mathbf{h}_{\mathrm{r,k}}^\mathsf{T},\mathbf{0}^\mathsf{T}]^\mathsf{T}$ for $k\in\mathcal{K}_1$ and $\bar{\mathbf{h}}_{\mathrm{r},k} = [\mathbf{0}^\mathsf{T},\mathbf{h}_{\mathrm{r,k}}^\mathsf{T}]^\mathsf{T}$ for $k\in\mathcal{K}_2$, and $P_\mathrm{A}$ denotes the power budget at active BD-RIS. Then the active BD-RIS operating power is modeled as $P_\mathrm{RIS}^\mathrm{tot} = P_\mathrm{c} + \frac{1}{\zeta}P_\mathrm{RIS}$, where $P_\mathrm{c}$ denotes the static circuit power and $\zeta\in(0,1]$ is the amplifier efficiency. Hence, the active BD-RIS energy consumption model is 
\begin{equation}
    E_\mathrm{RIS} = P_\mathrm{RIS}^\mathrm{tot}T_\mathrm{tx}, 
\end{equation}
where $T_\mathrm{tx} = \max_{k}~\alpha_k\frac{D_k}{R_k}$. 

The local execution energy of user $k$ is modeled and constrained as
\begin{equation}
E_k^{\mathrm{loc}}=(1-\alpha_k)\kappa_k D_k C_k (f_k^{\mathrm{loc}})^2\le E_k^\mathrm{max},
\end{equation}
where $\kappa_k$ is the effective switched-capacitance coefficient and $E_k^\mathrm{max}$ is the local energy budget for user $k$.

The offloading energy is
\begin{equation}
E_k^{\mathrm{off}}=\alpha_k \Big(p_k \frac{D_k}{R_k} + \kappa_\mathrm{edge}D_kC_k(f_k^\mathrm{edge})^2\Big),
\end{equation}
where $k_\mathrm{edge}$ is the edge computing-energy coefficient.


\subsection{Joint Optimization Problem}

Define the joint optimization variable set as
$\Omega \triangleq \{\boldsymbol{\alpha},\mathbf{p},\mathbf{f},\mathbf{W},\mathbf{\Theta}\}$,
where $\boldsymbol{\alpha} = [\alpha_1,\ldots,\alpha_K]^\mathrm{T}$ collects offloading decisions, $\mathbf{p} = [p_1,\ldots,p_K]^\mathsf{T}$ collects uplink transmit powers, $\mathbf{f}$ collects the local/edge computation allocation variables $\{f_k^\mathrm{edge}\}$, and $\mathbf{W} = [\mathbf{w}_1,\ldots,\mathbf{w}_K]$ denotes the combining matrix.
Then the joint offloading and resource-allocation problem to minimize the weighted system energy is formulated as
\begin{subequations}\label{eq:P1}
\begin{align}
\min_{\Omega}\quad
&
w_{\mathrm{loc}}\sum_{k=1}^{K}E_k^{\mathrm{loc}}
+
w_{\mathrm{off}}\sum_{k=1}^{K}E_k^{\mathrm{off}}
+
w_{\mathrm{RIS}}E_{\mathrm{RIS}}
\\
\text{s.t.}\quad
&
T_k\leq T_k^{\max},\quad \forall k,
\\
&E_k^\mathrm{loc}\le E_k^\mathrm{max}, \quad \forall k,\\
&
\alpha_k\in\{0,1\},\quad \forall k,
\\
&
p_{\min}\leq p_k\leq p_{\max},\quad \forall k,
\\
&
\sum_{k\in\mathcal{K}^{\mathrm{cpu,off}}} f_k^{\mathrm{edge}}\leq F_{\mathrm{cpu}}^{\mathrm{edge}},
\\
&
\sum_{k\in\mathcal{K}^{\mathrm{gpu,off}}} f_k^{\mathrm{edge}}\leq F_{\mathrm{gpu}}^{\mathrm{edge}},
\\
&
P_{\mathrm{RIS}}\leq P_A,\\
& \mathbf{\Theta}_{i,j}, \forall i,j\in\{1,2\}\text{~are~diagonal}, \mathbf{\Theta}_{1,2} = \mathbf{\Theta}_{2,1},
\end{align}
\end{subequations}
where $w_{\mathrm{loc}}$, $w_{\mathrm{off}}$, and $w_{\mathrm{RIS}}$ denote the weighting factors of local, offloading, and active BD-RIS energies, respectively, and $[p_\mathrm{min},p_\mathrm{max}]$ constraints the transmit power range.

Problem (\ref{eq:P1}) is a strongly coupled mixed-integer optimization  involving binary offloading, heterogeneous computation allocation, uplink power control, linear reception, and active BD-RIS configuration. Its nonconvexity and strong scenario dependence motivate the development of a learning-based slot-wise decision framework as will be detailed below.

\section{DSAC-T-Based Optimization Framework}

To enable efficient online decision making for problem ~(\ref{eq:P1}), we reformulate the active BD-RIS-assisted heterogeneous MEC process as a slot-wise learning problem. The original objective and constraints are incorporated through action projection and a penalty-based reward.


\begin{figure*}[t]
    \centering
    \includegraphics[width=0.95\textwidth]{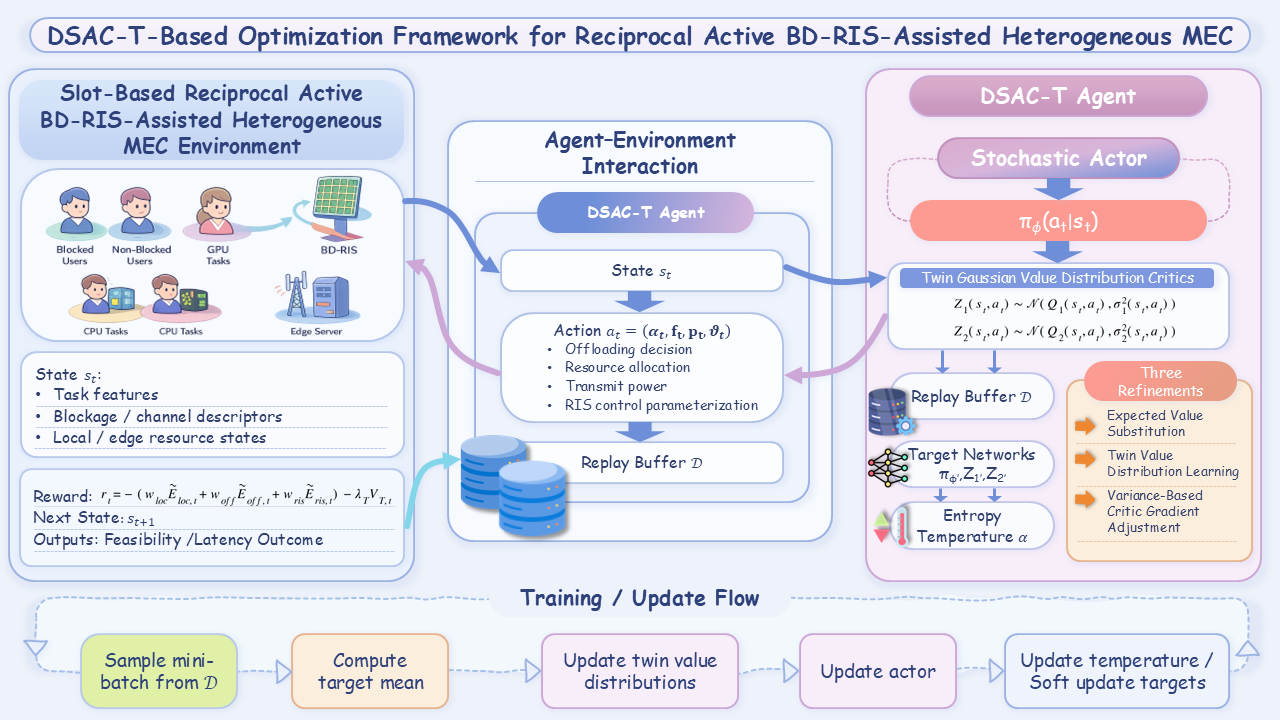}
    \caption{DSAC-T-based optimization framework for active BD-RIS-assisted heterogeneous MEC.}
    \label{fig:dsact_framework}
\end{figure*}

\subsection{MDP Reformulation}

The slot-wise decision process is modeled as
$\mathcal{M}=(\mathcal{S},\mathcal{A},\mathcal{P},\mathcal{R})$, where $\mathcal{S}$, $\mathcal{A}$, $\mathcal{P}$, and $\mathcal{R}$ denote the state space, action space, state transition probability, and the reward function, respectively. At slot $t$, the agent receives a state $s_t\in\mathcal{S}$ and selects an action $a_t\in\mathcal{A}$, and in return gains the next state $s_{t+1}\in\mathcal{S}$ and a reward $r_t\in\mathcal{R}$. Such a behavior of the agent is described as a policy $\pi(a_t~|~s_t)$ that maps from each state in $\mathcal{S}$ to a probability distribution over actions in $\mathcal{A}$.

In the considered scenario, the state $s_t\in\mathcal{S}$ at slot $t$ summarizes the task-side features $\{\mathbf{x}_{k,t}\}_{k=1}^K$, channel information $\{\mathbf{g}_{k,t}\}_{k=1}^K$, and local/edge resource conditions $\mathbf{q}_t$.

The action \(a_t\in\mathcal{A}\) contains the variables generated by the actor, including the offloading component, edge-computation allocation, uplink transmit powers, and the active BD-RIS control vector \(\boldsymbol{\vartheta}_t\). The raw action is processed by an environment-side mapping module: the offloading component is converted into binary decisions, the transmit powers are rescaled to the feasible range of offloaded users, and the edge-computation allocation is normalized within the corresponding CPU/GPU resource pools. The RIS-control component \(\boldsymbol{\vartheta}_t\) is mapped to a diagonal and reciprocal \(\mathbf{\Theta}_t\), followed by power projection if the active output-power constraint is violated. Given \(\mathbf{\Theta}_t\), the receive combiner \(\mathbf{W}_t\) is obtained by minimum mean square error (MMSE) reception.


Given $(s_t,a_t)$, the environment computes the communication/computation outcome and returns the next state by
\begin{equation}
s_{t+1}\sim \mathcal{P}(\cdot\mid s_t,a_t).
\end{equation}

The reward is designed as a penalty-based surrogate of problem (11), where the 
weighted energy objective is maximized in its negative form and latency violations 
are penalized explicitly as
\[
r_t=-
\left(
w_{\rm loc}\tilde{E}_{\rm loc,t}
+w_{\rm off}\tilde{E}_{\rm off,t}
+w_{\rm RIS}\tilde{E}_{\rm RIS,t}
\right)
-\lambda_T V_{T,t}.
\]
Here, $\tilde E_{\mathrm{loc},t}=\frac{\sum_k E_{k,t}^{\rm loc}}{E_{\mathrm{loc}}^{\mathrm{ref}}+\varepsilon}$,
$\tilde E_{\mathrm{off},t}=\frac{\sum_k E_{k,t}^{\rm off}}{E_{\mathrm{off}}^{\mathrm{ref}}+\varepsilon}$,
$\tilde E_{\mathrm{RIS},t}=\frac{E_{\mathrm{RIS},t}}{E_{\mathrm{RIS}}^{\mathrm{ref}}+\varepsilon}$ respectively denote the normalized slot-level local, offloading, and active BD-RIS energies,
with predefined normalization terms $E_{\mathrm{loc}}^{\mathrm{ref}}$, $E_{\mathrm{off}}^{\mathrm{ref}}$, $E_{\mathrm{RIS}}^{\mathrm{ref}}$, and $\varepsilon$. $V_{T,t}$ denotes the average relative latency violation, i.e.,
\begin{equation}
V_{T,t}
=
\frac{1}{K}\sum_{k=1}^{K}
\left[
\frac{T_{k,t}}{T_k^{\max}}-1
\right]_+,
\label{eq:vt}
\end{equation}
with $T_{k,t}$ being the end-to-end latency at slot $t$, and $\lambda_T$ denotes the penalty weight. 
To calculate the reward $r_t$, the digital combiner matrix $\mathbf{W}_t$ embedded in $E_{\mathrm{loc},t}$, $E_{\mathrm{off},t}$, and $E_{\mathrm{RIS},t}$ are directly optimized using typical MMSE receivers based on the effective channels as functions of $\boldsymbol{\vartheta}_t$.  
Hence, the reward directly penalizes weighted energy consumption and latency violation, while preserving the optimization intention of Section II.

\subsection{Why DSAC-T is Suitable for the Considered Problem}

The considered system exhibits sharp feasibility boundaries because several 
discrete and continuous decisions are coupled. A small change in the offloading 
decision may switch a user between local and edge execution, while a small change 
in transmit power or RIS coefficients may alter SINR, transmission latency, RIS 
output power, and latency feasibility. Moreover, CPU/GPU task heterogeneity and 
blocked/non-blocked channel conditions lead to highly non-uniform reward scales 
 across scenarios. Therefore, estimating only the expected return may be insufficient near 
feasibility boundaries, where overoptimistic value estimates can lead to unstable 
policy updates.


DSAC-T addresses this issue by modeling the state-action return of policy $\pi_\phi(a_t~|~s_t)$ using twin Gaussian value distributions
\begin{equation}
Z_{\theta_i}(s_t,a_t)\sim
\mathcal{N}\!\left(Q_{\theta_i}(s_t,a_t),\sigma_{\theta_i}^2(s_t,a_t)\right),
\label{eq:twin}
\end{equation}
where $\phi$ and $\theta_i$, $\forall i\in\{1,2\}$ are parameters.
This allows the critic to learn both the mean return and return uncertainty. Moreover, DSAC-T uses the smaller target mean
\begin{equation}
Q_{\theta_{\min}}(s_{t},a_{t})
=
\min_{i\in\{1,2\}}
Q_{\theta_i}(s_{t},a_{t}),
\label{eq:q_target_min}
\end{equation}
to construct and update the target. 

The twin-distribution design and variance-aware update thus reduce overestimation bias and improve robustness under heterogeneous energy-latency rewards, making DSAC-T suitable for problem~(\ref{eq:P1}).

\subsection{DSAC-T Based Learning Algorithm}

Based on the MDP reformulation, we adopt a DSAC-T based learning algorithm including the following two steps.

\textit{Step 1: Sampling.}  
At each slot $t$, the agent observes the current state $s_t$, samples an action $a_t\sim\pi_\phi(\cdot\mid s_t)$, receives the reward $r_t$ and the next state $s_{t+1}$ from the environment, and stores the transition in the replay buffer:
\begin{equation}
\mathcal{D}
\leftarrow
\mathcal{D}\cup\{(s_t,a_t,r_t,s_{t+1})\}.
\end{equation}

\textit{Step 2: Updating.}
Given a transition $(s_t,a_t,r_t,s_{t+1})$ sampled from the replay buffer $\mathcal{D}$, the target action is sampled by
$a_{t+1}\sim \pi_{\phi'}(\cdot\mid s_{t+1})$,
where $\phi'$ denotes a parameter for target networks. 
Define the target self-consistency operator as 
\begin{equation}
\mathcal{T}_{\phi',i}
=
r_t+\gamma\left(
Z_i(s_{t+1},a_{t+1})
-\alpha\log\pi_{\phi'}(a_{t+1}\mid s_{t+1})
\right),
\end{equation}
where $\gamma\in(0,1)$ is a discount factor and $\alpha$ is the temperature coefficient. 
Then the twin value distributions $\theta_1$ and $\theta_2$ are updated by minimizing 
\begin{equation}
\mathcal{L}_{\mathcal{Z}} = \omega_i \mathbb{E}\{F(\mathcal{Y}_{\theta'_i}(\cdot~|~s_t,a_t),\mathcal{Z}_{\theta_i}(\cdot~|~s_t,a_t))\},
\end{equation}
where $\omega_i =\mathbb{E}\{\sigma_{\theta_i}^2(s_t,a_t)\}$ is a gradient scaling weight, $F(\cdot,\cdot)$ denotes a distance measurement function, $\mathcal{Y}_{\theta'_i}(\cdot~|~s_t,a_t)$ denotes the distribution of $\mathcal{T}_{\phi',i}$ with $\theta_1'$ and $\theta_2'$ being parameters for target networks related to the distributions of $Z_1(s_{t+1},a_{t+1})$ and $Z_2(s_{t+1},a_{t+1})$. 
The actor $\phi$ is updated by minimizing 
\begin{equation}
\mathcal{L}_{\pi}
=
\mathbb{E}
\{
\alpha \log \pi_\phi(a_t\mid s_t)
- Q_{\theta_{\min}}(s_t,a_t)\}.
\end{equation}
The entropy temperature is updated by minimizing 
\begin{equation}
\mathcal{L}_{\alpha}
=
\mathbb{E}
\{
-\alpha\big(\log \pi_\phi(a_t\mid s_t)+\mathcal{H}_{\mathrm{tar}}\big)
\},
\end{equation}
where $\mathcal{H}_\mathrm{tar}$ denotes the expected entropy.
The update of twin value distributions, actor, and temperature is practically done by the gradient descent, i.e.,
\begin{equation}
\theta_i
\leftarrow
\theta_i-\eta_\mathcal{Z}\nabla_{\theta_i}\mathcal{L}_{\mathcal{Z}},
\phi
\leftarrow
\phi-\eta_\pi\nabla_\phi\mathcal{L}_{\pi},
\alpha
\leftarrow
\alpha-\eta_\alpha\nabla_\alpha\mathcal{L}_{\alpha},
\end{equation}
where $\eta_\mathcal{Z}$, $\eta_\pi$, and $\eta_\alpha$ are learning rates and $\nabla_{\theta_i}\mathcal{L}_{\mathcal{Z}}$, $\nabla_{\phi}\mathcal{L}_\pi$, and $\nabla_\alpha\mathcal{L}_\alpha$ are first-order derivatives.  
Finally, the target networks are softly updated according to
\begin{equation}
\theta_i'
\leftarrow
\tau\theta_i+(1-\tau)\theta'_i,~~ \phi'
\leftarrow
\tau\phi+(1-\tau)\phi',
\end{equation}
where $\tau$ is the soft-update coefficient.

The above two steps iterate with each other until convergence. 
To summarize the interaction between the active BD-RIS-assisted heterogeneous MEC system, the DSAC-T agent, as well as the corresponding replay-based learning pipeline, we illustrate the overall framework in Fig.~\ref{fig:dsact_framework}.

\section{Performance Evaluation} 
We consider a reciprocal active BD-RIS-assisted heterogeneous MEC system with $K=10$ users  (including $4$ deeply blocked users), one BS with $N=8$ antennas, and one hybrid mode active BD-RIS with $M=128$ cells. The system bandwidth is $B=16$ MHz. The direct link of blocked users is modeled as a severely attenuated residual path with an additional attenuation factor of $0.05$. The wireless channels follow distance-dependent Rician fading. The direct user--BS distance is sampled from $[20,120]$ m, the user--RIS and RIS--BS distances from $[5,20]$ m, and the blocked-user--RIS distance from $[5,15]$ m. The path-loss exponents of the direct, user--RIS, and RIS--BS links are $3.5$, $2.2$, and $2.2$, with large-scale scaling factors $0.8$, $1.0$, and $1.1$, respectively. The corresponding Rician factors are $2$ dB, $6$ dB, and $10$ dB. The BS and RIS noise powers are $\sigma_\mathrm{B}^2=10^{-9}$ and $\sigma_\mathrm{I}^2=1.5\times10^{-9}$, respectively. 
The uplink transmit power ranges from $0.01$ W to $0.1$ W for offloaded users. The active BD-RIS has static power $0.5$ W, amplifier efficiency $0.4$, and active output-power budget $0.01$ W. The maximum local CPU frequency is $1.3$ GHz, while the edge CPU and GPU pools are $24$ GHz and $48$ GHz, respectively. Each evaluation scenario contains $5$ CPU-type and $5$ GPU-type tasks. CPU-type tasks have data sizes in $[8\times10^4,1.8\times10^5]$ bits, computation intensities in $[400,650]$ cycles/bit, and latency deadlines in $[0.35,0.8]$ s. GPU-type tasks have data sizes in $[1.98\times10^5,4.5\times10^5]$ bits, computation intensities in $[900,1300]$ cycles/bit, and latency deadlines in $[0.198,0.462]$ s. The reward weights are set to $w_{\mathrm{loc}}=0.5$, $w_{\mathrm{off}}=2.0$, $w_{\mathrm{RIS}}=0.5$, and $\lambda_T=1.0$, emphasizing offloading-related energy while retaining an explicit penalty for active BD-RIS overhead induced by cross-sector support. 

\begin{figure}[t] 
\centering 
\includegraphics[width=0.45\textwidth]{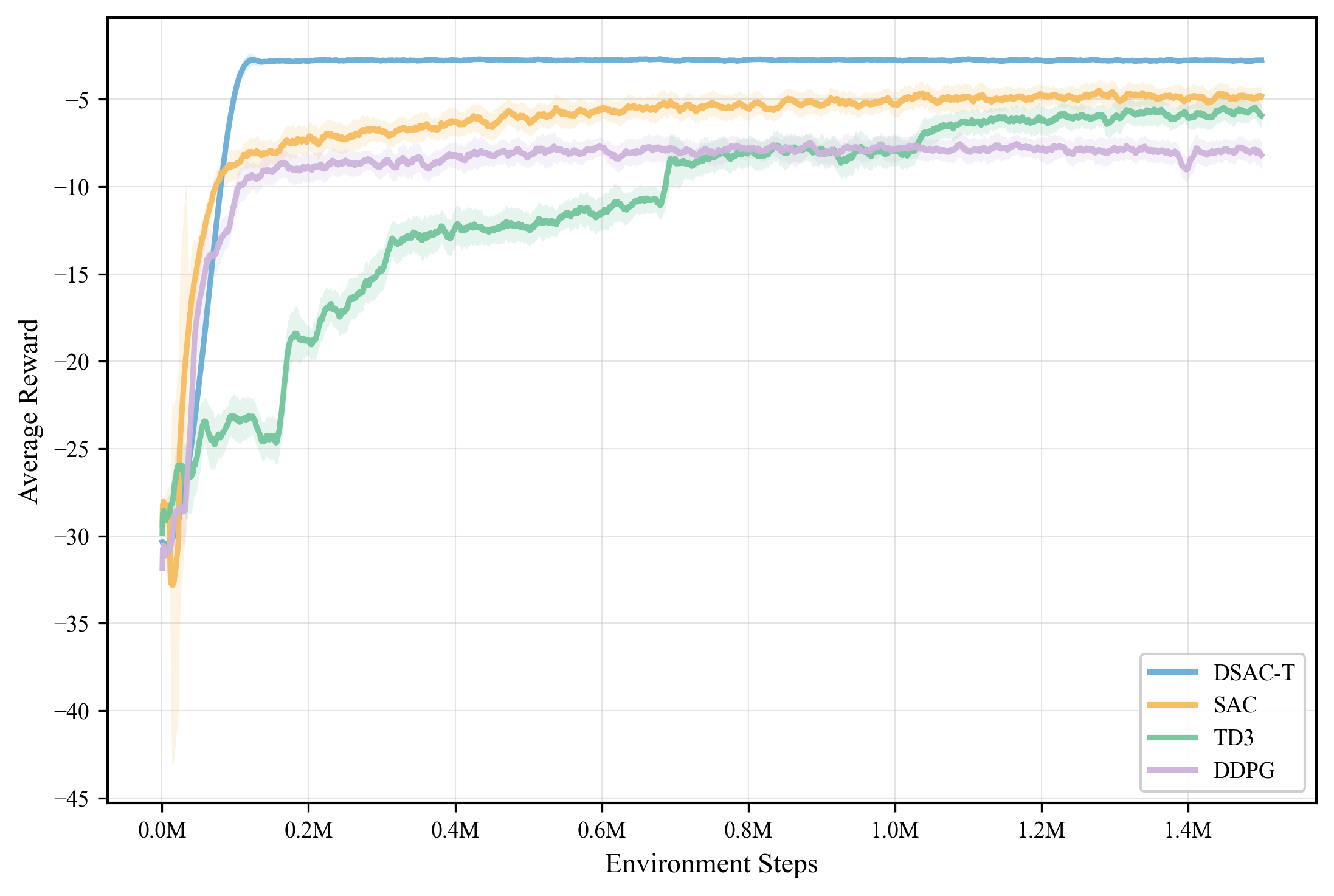} 
\caption{Training convergence curves of different DRL methods.} 
\label{fig:learning_curve} 
\end{figure} 

Fig.~\ref{fig:learning_curve} compares the training convergence of different DRL methods. DSAC-T converges rapidly and maintains the highest reward after convergence. SAC also converges quickly but stabilizes at a lower level, while Twin Delayed Deep Deterministic Policy Gradient (TD3) improves more slowly and Deep Deterministic Policy Gradient (DDPG) achieves lower final performance. This shows that the distributional value learning and refinement mechanisms in DSAC-T improve convergence quality and policy stability. 

\begin{table}[t] 
\centering 
\caption{Performance-Complexity Comparison over 300 Evaluation Scenarios} 
\label{tab:performance_runtime} 
\footnotesize \begin{tabular}{lccc} 
\toprule Method & Reward & Feasibility Ratio & Runtime$^{\ddagger}$ (s/scenario) 
\\ \midrule 
DSAC-T & -2.828 & 81.67\% & 0.0267 \\ DDPG & -5.899 & 64.33\% & 0.0494 \\ SAC & -4.068 & 71.33\% & 0.0516 \\ TD3 & -5.135 & 70.67\% & 0.0533 \\ AO--SCA & -8.393 & 66.67\% & 41.865 \\ 
\bottomrule \end{tabular} 
\vspace{0.5mm} 
\begin{flushleft} 
\scriptsize $^\ddagger$The average online decision time per scenario; training time is excluded. 
\end{flushleft} 
\end{table} 
Table~\ref{tab:performance_runtime} reports the final performance and online decision complexity. Since the reward is defined as the negative energy-latency penalty, a larger value is better. DSAC-T achieves the best reward and the highest feasibility ratio, improving feasibility by $10.34$ percentage points over SAC and $11.00$ percentage points over TD3. It also requires only $0.0267$ s per scenario, whereas the alternating optimization (AO) method based on successive convex approximation (SCA) requires $41.865$ s due to iterative per-instance optimization. These results demonstrate that DSAC-T achieves the best performance-complexity tradeoff among the compared methods. 

\begin{figure}[t] 
\centering 
\includegraphics[width=0.48\textwidth]{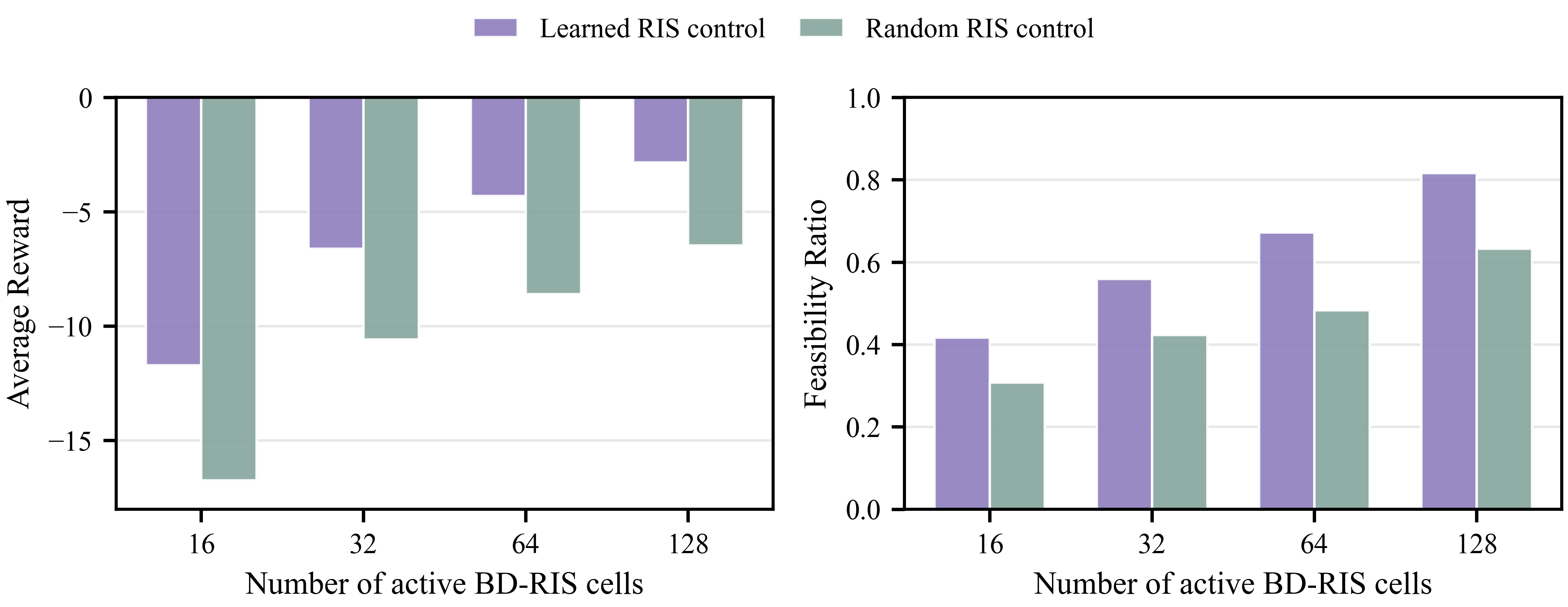} 
\caption{Impact of the number of active BD-RIS cells on reward and feasibility.} 
\label{fig:ris_elements} 
\end{figure} 

Fig.~\ref{fig:ris_elements} evaluates the impact of the number of active BD-RIS cells. Increasing the number of cells improves both reward and feasibility ratio, since a larger active BD-RIS provides more wave manipulation degrees of freedom. In addition, learned RIS control consistently outperforms random RIS control under all cell number settings, confirming that explicit RIS coefficient optimization is necessary and that the gain does not simply come from increasing the aperture size.

\section{Conclusion}

This paper studied energy-aware offloading and resource allocation for reciprocal active BD-RIS-assisted heterogeneous MEC. A physically grounded model was established to capture asymmetric blockage, residual direct-link attenuation, active surface power consumption, amplified noise, reciprocal cross-sector leakage, and decoupled CPU/GPU edge resources. The resulting mixed integer and strongly coupled problem was reformulated as a 
slot-wise learning-based decision problem, where the original objective and constraints were handled through action processing and a penalty-based reward. 
A DSAC-T-based framework was then developed with real-valued action parameterization for joint offloading, computation allocation, power control, and active BD-RIS control. Simulation results showed that DSAC-T achieved 
the best energy-latency reward and the highest feasibility ratio among the compared methods, while maintaining low online decision latency. These results validate the effectiveness of distributional value learning for feasibility-sensitive and reward-heterogeneous active BD-RIS-assisted MEC optimization.

\section*{Acknowledgment}

This work is funded by the National Natural Science Foundation of China (grant no. 62501509) and the Natural Science Foundation of Guangdong Province (grant no. 2026A1515011048).

\bibliographystyle{IEEEtran}
\bibliography{references}

@ARTICLE{Mao2017Survey,
  author={Mao, Yuyi and You, Changsheng and Zhang, Jun and Huang, Kaibin and Letaief, Khaled B.},
  journal={IEEE Communications Surveys \& Tutorials}, 
  title={A Survey on Mobile Edge Computing: The Communication Perspective}, 
  year={2017},
  volume={19},
  number={4},
  pages={2322-2358},
  doi={10.1109/COMST.2017.2745201}}

@article{liu2021reconfigurable,
  title={Reconfigurable intelligent surfaces: Principles and opportunities},
  author={Liu, Yuanwei and Liu, Xiao and Mu, Xidong and Hou, Tianwei and Xu, Jiaqi and Di Renzo, Marco and Al-Dhahir, Naofal},
  journal={IEEE Communications Surveys \& Tutorials},
  volume={23},
  number={3},
  pages={1546--1577},
  year={2021},
  publisher={IEEE}
}

@ARTICLE{Li2026Tutorial,
  author={Li, Hongyu and Nerini, Matteo and Shen, Shanpu and Clerckx, Bruno},
  journal={IEEE Communications Surveys \& Tutorials}, 
  title={A Tutorial on Beyond-Diagonal Reconfigurable Intelligent Surfaces: Modeling, Architectures, System Design and Optimization, and Applications}, 
  year={2026},
  volume={28},
  number={},
  pages={4086-4126},
  doi={10.1109/COMST.2025.3647003}}

@article{Liu2021STAR360Coverage,
  author    = {Yuanwei Liu and Xidong Mu and Junqing Xu and Robert Schober and Yue Gao and H. Vincent Poor and Lajos Hanzo},
  title     = {{STAR}: Simultaneous Transmission and Reflection for 360° Coverage by Intelligent Surfaces},
  journal   = {IEEE Wireless Communications},
  volume    = {28},
  number    = {6},
  pages     = {102--109},
  year      = {2021},
  doi       = {10.1109/MWC.001.2100191}
}

@article{Hu2021RISMECOptimizationLearning,
  author    = {Xiaoyan Hu and Christos Masouros and Kai-Kit Wong},
  title     = {Reconfigurable Intelligent Surface Aided Mobile Edge Computing: From Optimization-Based to Location-Only Learning-Based Solutions},
  journal   = {IEEE Transactions on Communications},
  volume    = {69},
  number    = {6},
  pages     = {3709--3725},
  year      = {2021},
  doi       = {10.1109/TCOMM.2021.3066495}
}

@ARTICLE{Qin2025Joint,
  author={Qin, Xintong and Yu, Wenjuan and Ni, Qiang and Song, Zhengyu and Hou, Tianwei and Sun, Xin},
  journal={IEEE Communications Letters}, 
  title={Joint Resource Allocation and Beamforming Design for {BD-RIS}-Assisted Wireless-Powered Cooperative Mobile Edge Computing}, 
  year={2025},
  volume={29},
  number={5},
  pages={1042-1046},
  doi={10.1109/LCOMM.2025.3552901}}

@article{Long2021ActiveRISAided,
  author    = {Ruizhe Long and Ying-Chang Liang and Yiyang Pei and Erik G. Larsson},
  title     = {Active Reconfigurable Intelligent Surface-Aided Wireless Communications},
  journal   = {IEEE Transactions on Wireless Communications},
  volume    = {20},
  number    = {8},
  pages     = {4962--4975},
  year      = {2021},
  doi       = {10.1109/TWC.2021.3064024}
}

@ARTICLE{Xu2023Active,
  author={Xu, Jiaqi and Zuo, Jiakuo and Zhou, Joey Tianyi and Liu, Yuanwei},
  journal={IEEE Communications Letters}, 
  title={Active Simultaneously Transmitting and Reflecting {(STAR)-RISs}: Modeling and Analysis}, 
  year={2023},
  volume={27},
  number={9},
  pages={2466-2470},
  doi={10.1109/LCOMM.2023.3289066}}

@article{shen2026active,
  title={Active Beyond-Diagonal Reconfigurable Intelligent Surfaces: Modeling, Architecture Design, and Optimization},
  author={Shen, Shanpu and Li, Hongyu and Nerini, Matteo and Wu, Qingqing and Clerckx, Bruno},
  journal={arXiv:2603.13861},
  year={2026}
}

@article{liu2026active,
  title={Active Beyond-Diagonal Reconfigurable Intelligent Surface with Hybrid Transmitting and Reflecting Mode},
  author={Liu, Fu and Li, Hongyu and Shen, Shanpu},
  journal={arXiv:2604.13570},
  year={2026}
}

@article{aung2025active,
  title={Active {STAR-RIS} empowered edge system for enhanced energy efficiency and task management},
  author={Aung, Pyae Sone and Kim, Kitae and Tun, Yan Kyaw and Huh, Eui-Nam and Han, Zhu and Hong, Choong Seon},
  journal={IEEE Transactions on Mobile Computing},
  year={2025},
  publisher={IEEE}
}

@article{qin2025resource,
  title={Resource Allocation and Beamforming Design for Active {STAR-RIS}-Assisted Wireless-Powered {MEC}},
  author={Qin, Xintong and Yu, Wenjuan and Ni, Qiang and Song, Zhengyu and Hou, Tianwei and Wang, Jun and Sun, Xin},
  journal={IEEE Internet of Things Journal},
  year={2025},
  publisher={IEEE}
}

@article{Huang2020DRLOnlineOffloadingMEC,
  author    = {L. Huang and others},
  title     = {Deep Reinforcement Learning for Online Computation Offloading in Wireless Powered Mobile-Edge Computing Networks},
  journal   = {IEEE Transactions on Mobile Computing},
  year      = {2020},
  doi       = {10.1109/TMC.2019.2928811}
}

@article{Peng2023RobustDRLForEHRIS,
  author    = {H. Peng and X. Wang},
  title     = {Energy Harvesting Reconfigurable Intelligent Surface for {UAV} Communication Based on Robust Deep Reinforcement Learning},
  journal   = {IEEE Transactions on Wireless Communications},
  year      = {2023},
  doi       = {10.1109/TWC.2023.3241168}
}

@ARTICLE{Duan2025DSACT,
author={Duan, Jingliang and Wang, Wenxuan and Xiao, Liming and Gao, Jiaxin and Li, Shengbo Eben and Liu, Chang and Zhang, Ya-Qin and Cheng, Bo and Li, Keqiang},
journal={IEEE Transactions on Pattern Analysis \& Machine Intelligence},
title={Distributional Soft Actor-Critic With Three Refinements},
year={2025},
volume={47},
number={05},
pages={3935-3946},
doi={10.1109/TPAMI.2025.3537087}
}

\end{document}